\documentclass[article,preprint,groupedaddress]{revtex4}
\usepackage{epsfig}
\usepackage{graphicx}

\begin{document}

\title{Einstein-Yang-Mills Solitons: The Role of Gravity}
\author{Shahar Hod}
\address{The Ruppin Academic Center, Emeq Hefer 40250, Israel}
\address{and}
\address{The Hadassah Institute, Jerusalem 91010, Israel}
\date{\today}

\begin{abstract}

\ \ \ The canonical Bartnik-McKinnon solitons are regular solutions
of the coupled Einstein-Yang-Mills system in which gravity may
balance the repulsive nature of the Yang-Mills field. We examine the
role played by gravity in balancing the system and determine its
strength. In particular, we obtain an analytic lower bound on the
fundamental mass-to-radius ratio, ${\max}_r\{2m(r)/r\}>2/3$, which
is a necessary condition for the existence of globally regular
Einstein-Yang-Mills solitons. Our analytical results are in accord
with numerical calculations.
\end{abstract}
\bigskip
\maketitle

The wide interest in the theory of gravitating solitons and hairy
black holes started after the (numerical) discovery of globally
regular particle-like solutions of the coupled Einstein-Yang-Mills
(EYM) system \cite{BarMic,BizCol}. This was rather unexpected, since
it is well known that when taken apart, neither the vacuum Einstein
equations nor the Yang-Mills (YM) equations have nontrivial static
globally regular solutions. For the pure YM theory in flat space,
this was proved in \cite{Des,Cole}. The corresponding result for
vacuum Einstein gravity is Lichnerowicz's theorem \cite{Lich}.
Indeed, the unique static spherically symmetric solution of pure
Einstein gravity is the celebrated Schwarzschild metric, which is
singular at the origin.

The EYM solitons (also known as Bartnik-McKinnon solitons
\cite{BarMic}) can be thought of as equilibrium states of a pair of
physical fields \cite{GalSum}, one of which is repulsive (the YM
field) while the other one is attractive (gravity). It turns out
that the YM repulsive force can balance gravitational attraction and
prevent the formation of singularities in spacetime. From a
mathematical point of view, it is the nonlinearity of the YM
equations which may preclude spacetime singularities.

The fact that only the coupled EYM system can have regular solitons
(while the pure YM equations can have no static regular solutions)
raises the fundamental question: how strong must gravity be in order
to balance the repulsive nature of the YM field? Here we shall give
an analytic answer to this question by proving the existence of a
lower bound on the fundamental mass-to-radius ratio,
${\max}_r\{2m(r)/r\}$. This quantity is central for determining the
spacetime geometry of the solution and the strength of the
gravitational interaction.

The line element of a static spherically symmetric spacetime may
take the following form in Schwarzschild coordinates \cite{Nun}

\begin{equation}\label{Eq1}
ds^2=-e^{-2\delta}\mu dt^2 +\mu^{-1}dr^2+r^2(d\theta^2 +\sin^2\theta
d\phi^2)\  ,
\end{equation}
where the metric functions $\delta$ and $\mu=1-2m(r)/r$ depend only
on the Schwarzschild radius $r$. Here $m(r)$ is the mass contained
within a sphere of radius $r$. (We use gravitational units in which
$G=c=1$.) Asymptotic flatness requires that as $r \to \infty$,

\begin{equation}\label{Eq2}
\mu(r) \to 1\ \ \ and\ \ \ \ \delta(r) \to 0\  ,
\end{equation}
and a regular center requires

\begin{equation}\label{Eq3}
\mu(r) =1+O(r^2)\  .
\end{equation}

The Einstein equations, $G^{\mu}_{\nu}=8\pi T^{\mu}_{\nu}$, reads

\begin{equation}\label{Eq4}
\mu^{'}=8\pi rT^{t}_{t}+(1-\mu)/r\  ,
\end{equation}

\begin{equation}\label{Eq5}
\delta^{'}=4\pi r(T^{t}_{t}-T^{r}_{r})/\mu\  ,
\end{equation}
where the prime stands for differentiation with respect to $r$. The
conservation equation, $T^{\mu}_{\nu ;\mu}=0$, has only one
nontrivial component \cite{Nun}

\begin{equation}\label{Eq6}
T^{\mu}_{r ;\mu}=0\  .
\end{equation}

Substituting Eqs. (\ref{Eq4}) and (\ref{Eq5}) in Eq. (\ref{Eq6}),
one obtains

\begin{equation}\label{Eq7}
(e^{-\delta} r^4 T^{r}_{r})^{'}= {{e^{-\delta}r^3} \over
{2\mu}}[(3\mu-1) (T^{r}_{r}-T^{t}_{t})+2\mu T]\  ,
\end{equation}
where $T$ is the trace of the energy momentum tensor. Below we shall
concentrate on the function ${\cal E}(r) \equiv e^{-\delta} r^4
T^{r}_{r}$.

Our main focus here is on the canonical EYM solitons. However, our
analytical results would also be valid for any Einstein-matter
theory in which the matter fields satisfy the following energy
conditions:
\begin{itemize}
\item{The weak energy condition (WEC). This means that the energy
density, $\rho \equiv -T^{t}_{t}$, is positive semidefinite and that
it bounds the pressures, in particular, $|T^{r}_{r}| \leq \rho$.
This implies the inequality $T^{r}_{r}-T^{t}_{t} \geq 0$.}
\item{The energy density $\rho$ goes to zero faster than $r^{-4}$. This
requirement is the natural way to impose the condition that there
are no extra conserved charges (besides the ADM mass) defined at
asymptotic infinity associated with the matter fields \cite{Nun}.
(We recall that the charges defined at spatial infinity, like the
electric charge of the Reissner-Nordstr\"om solution in
Einstein-Maxwell theory, are associated with the $\rho \sim r^{-4}$
asymptotic behavior.)}
\item{The trace of the energy momentum tensor is nonnegative, $T \geq 0$.}
\end{itemize}
We point out that the EYM system satisfies these energy conditions,
in particular, $\rho(r \to \infty) \sim r^{-6}$ and $T=0$.

Taking cognizance of Eq. (\ref{Eq7}) together with the boundary
conditions, Eqs. (\ref{Eq2})-(\ref{Eq3}), and the above energy
conditions one finds that as $r \to 0^{+}$,

\begin{equation}\label{Eq8}
{\cal E}\to 0^{+}\ \ \ and\ \ \ \ {\cal E}^{'} \geq 0\  ,
\end{equation}
and that as $r \to \infty$,

\begin{equation}\label{Eq9}
{\cal E}\to 0^{-}\ \ \ and\ \ \ \ {\cal E}^{'} \geq 0\  ,
\end{equation}
Equations (\ref{Eq8}) and (\ref{Eq9}) imply that there is a finite
interval, $r_0 \leq r \leq r_1$, in which ${\cal E}^{'}(r)<0$. (The
function ${\cal E}$ switches signs from positive values to negative
values within this interval). For ${\cal E}^{'}(r)$ to be negative,
we must have $3\mu(r) -1<0$ in this interval. This follows from the
WEC and the assumption that $T \geq 0$. This, in turn, yields a
lower bound on the maximal mass-to-radius ratio of the regular
soliton:

\begin{equation}\label{Eq10}
{\max}_r \Big\{{{2m(r)} \over r}\Big\} > {2\over 3}\  .
\end{equation}

The dimensionless quantity $2m(r)/r$ is fundamental for determining
the spacetime geometry and the strength of the gravitational
interaction. The analytically derived inequality therefore sets a
lower bound on the strength of gravity which is required in order to
stabilize the repulsive nature of the YM field.

We would like to point out that {\it upper} bounds on the ratio
${\max}_r\{2m(r)/r\}$ are well known in the literature, the famous
of all is the Buchdahl inequality, ${\max}_r\{2m(r)/r\}>8/9$
\cite{Buch,Hak}. However, this upper bound has been derived for
matter fields which satisfy the conditions $T \leq 0$ and
$T^{r}_{r}\geq 0$, and it should be emphasized that this assumption
is violated by the YM system considered here (we have proved that
$T^{r}_{r}$ must switch signs for the EYM system, and this
analytical prediction is in accord with numerical calculations, see
\cite{GalSum}.) To our best knowledge, the inequality (\ref{Eq10})
is the first {\it lower} bound derived on the fundamental ratio
${\max}_r\{2m(r)/r\}$.

Let us confirm the validity of the analytical relation, Eq.
(\ref{Eq10}), with the help of available numerical data. It turns
out \cite{Breit} that ${\max}_r\{2m/r\}=0.76>2/3$ for the canonical
$n=1$ EYM soliton (the positive integer $n$ is the number of nodes
of the matter fields.) In addition, the function ${\max}_r\{2m/r\}$
increases as $n$ increases. These numerical results are obviously in
accord with our analytical bound, Eq. (\ref{Eq10}).

{\it Summary.---} The nonlinearity of the YM equations may preclude
spacetime singularities, and thus allows the existence of globally
regular solitons solutions of the coupled EYM system. On the other
hand, this nonlinearity has restricted most former studies of the
EYM equations to the numerical regime. It is therefore highly
important to obtain some {\it analytical} insights about this
nonlinear system. This was the main purpose of our analysis.

It is well known that the pure YM equations have no regular static
solutions due to the repulsive nature of the YM field. It is the
attractive nature of gravity that allows the existence of regular
EYM solitons. One would therefore like to examine the role played by
gravity in balancing the coupled system. Here we have proved
analytically that the fundamental mass-to-radius ratio must be
bounded according to ${\max}_r\{2m(r)/r\}>2/3$ in order to balance
the system and to allow the existence of globally regular solitons.
We find it remarkable that such a simple and transparent relation
emerged out of the highly nonlinear equations.

\bigskip
\noindent{\bf ACKNOWLEDGMENTS}
\bigskip

This research is supported by the Meltzer Science Foundation. I
thank Uri Keshet for stimulating discussions.


\begin{thebibliography}{99}

\bibitem{BarMic} R. Bartnik and J. McKinnon, Phys. Rev. Lett. {\bf
61}, 141 (1998).

\bibitem{BizCol} The discovery of the EYM regular solitons has subsequently led to the discovery of
the EYM hairy black holes, see P. Bizo\'n, Phys. Rev. Lett {\bf 64},
2844 (1990); M. S. Volkov and D. V. Gal'tsov, Sov. J. Nucl. Phys.
{\bf 51}, 1171 (1990).

\bibitem{Des} S. Deser, Phys. Lett. B {\bf 64}, 463 (1976).

\bibitem{Cole} S. Coleman, in {\it New Phenomena in Subnuclear
Physics}, edited by A. Zichichi (Plenum, New York, 1975).

\bibitem{Lich} A. Lichnerowicz, in {\it Les Theories Relativistes de
la Gravitation} (Masson, Paris, 1955).

\bibitem{GalSum} M. S. Volkov and D. V. Gal'tsov, Physics Reports {\bf
319}, 1 (1999).

\bibitem{Nun} D. N\'u\~nez, H. Quevedo, and D. Sudarsky, Phys. Rev. Lett. {\bf 76}, 571 (1996).

\bibitem{Buch} H. A. Buchdahl, Phys. Rev. {\bf 116}, 1027 (1959).

\bibitem{Hak} H. Andr\'easson, arXiv:gr-qc/0702137.

\bibitem{Breit} P. Breitenlohner, P. Forg\'acs, and D. Maison, Comm.
Math. Phys. {\bf 163}, 141 (1994).

\end{thebibliography}
\end{document}